\documentclass[twocolumn, superscriptaddress, preview, amsmath, amssymb, aps, letterdraft, notitlepage]{revtex4-1}

\usepackage{graphicx}
\usepackage{dcolumn}
\usepackage{bm}

\usepackage[colorlinks=true,linkcolor=blue,citecolor=blue,urlcolor=blue]{hyperref}
\usepackage[caption=false]{subfig}

\usepackage{cleveref}

\usepackage{braket}
\usepackage{ulem}

\begin{document}
\title{Stability and Dynamics of Many-Body Localized Systems Coupled to Small Bath}

\author{Shao-Hen Chiew}

\affiliation{Department of Physics, Faculty of Science
National University of Singapore
Blk S12 Level 2, Science Drive 3
Singapore 117551}
\affiliation{Centre for Quantum Technologies, National University of Singapore}
\author{Jiangbin Gong}
\affiliation{Department of Physics, Faculty of Science
National University of Singapore
Blk S12 Level 2, Science Drive 3
Singapore 117551}
\author{Leong-Chuan Kwek}
\affiliation{Centre for Quantum Technologies, National University of Singapore} 
\affiliation{MajuLab, CNRS-UNS-NUS-NTU International Joint Research Unit, Singapore UMI 3654, Singapore}
\affiliation{National Institute of Education, Nanyang Technological University, Singapore 637616, Singapore}
\affiliation{Quantum Science and Engineering Centre (QSec), Nanyang Technological University, Singapore}
\author{Chee-Kong Lee}
\email{cheekonglee@tencent.com}
\affiliation{Tencent America, Palo Alto, CA 94306, United States}



\begin{abstract}
It is known that strong disorder in closed quantum systems leads to many-body localization (MBL), and that this quantum phase can be destroyed by coupling to an infinitely large Markovian environment. However, the stability of the MBL phase is less clear when the system and environment are of finite and comparable size.
Here, we study the stability and eventual localization properties of a disordered Heisenberg spin chain coupled to a finite environment, and extensively explore the effects of environment disorder, geometry, initial state and system-bath coupling strength.
By studying the non-equilibrium dynamics and the eventual steady-state properties of different initial states, our numerical results indicate that in most cases, the system retains its localization properties despite the coupling to the finite environment, albeit to a reduced extent. 
However, in cases where the system and environment is strongly coupled in the ladder configuration, the eventual localization properties are highly dependent on the initial state, and could lead to either thermalization or localization.
\end{abstract}

\maketitle

\section{Introduction}

Dynamics in ergodic systems typically lead to thermalization, in which the system converges to a thermodynamic equilibrium state independent of its initial state \cite{deutsch1991quantum, srednicki1994chaos, rigol2008thermalization}. Yet, from the point of view of quantum information processing, understanding conditions in which thermalization fails - thus resulting in the retention of information - is attractive. Interacting many-body quantum systems that are able to retain information, said to exhibit many-body localization (MBL) \cite{nandkishore2015many1, alet2018many, pal2010many,
abanin2019manybody}, allow us to contrive dynamics which maintain quantum information in the presence of interactions. It is now known that strong disorder can drive closed ergodic systems into localization, a fact supported by large bodies of theoretical \cite{serbyn2013local, huse2014phenomenology, imbrie2016many}, experimental \cite{schreiber2015observation, smith2016many, choi2016exploring, bordia2017probing, roushan2017spectroscopic, xu2018emulating} and numerical \cite{oganesyan2007localization, pal2010many, luitz2015many} work.

In practical situations, quantum systems are not ideally isolated and may be coupled to an ergodic environment to varying extents. Even in the ideal isolated situation, rare regions of low disorder that form amidst a disordered system effectively act as ergodic subsystems with the potential to thermalize the larger system, depending on dimensionality and the nature of the couplings \cite{agarwal2017rare, de2017many, thiery2018many, de2017stability}. An understanding on the resulting fate of such interactions between localized and ergodic phases is therefore important to establish the stability and robustness of MBL, and whether it can survive in higher dimensions \cite{hyatt2017many, marino2018many, nandkishore2015many, doggen2020slow}. This is also experimentally interesting \cite{rubioabadal2019many, bordia2017probing}, especially in higher dimensional systems which are still inaccessible to numerical simulations. 

While one can model the system-bath interaction with quantum master equations such as the Lindblad master equation, the number of degrees of freedom of the bath is assumed to be large compared to that of the system, and with this approach the system is expected to thermalize at long times \cite{levi2016robustness, fischer2016dynamics, medvedyeva2016influence,wei2018exploring}. However, the situation is less clear if the number of degrees of freedom of the environment is comparable to that of the system, in which backaction and proximity effects can be significant. Under certain conditions, it has been shown that these effects can prevent quantum systems from thermalization \cite{li2015many, nandkishore2015many, wybo2020entanglement,huse2013localization}, which can be a desirable scenario for quantum information processing. Moreover, there has been less emphasis on understanding the effects of such couplings from a dynamical perspective, and its relation to results obtained from the entire eigenspectrum of MBL systems, which is important in understanding many existing experimental results \cite{schreiber2015observation, smith2016many, choi2016exploring, bordia2017probing, xu2018emulating}

In this article, we focus on the situation where the number of degrees of freedom of the environment is comparable to that of the system, and ask whether such couplings preserve or destroy the localization properties of the system or the environment. Does a localised system lose its localisation properties when coupled to an ergodic bath, or is the bath localised instead? Do they retain their initial localisation properties? If so, to what extent? 

\begin{figure}[h]
\includegraphics[scale=0.3]{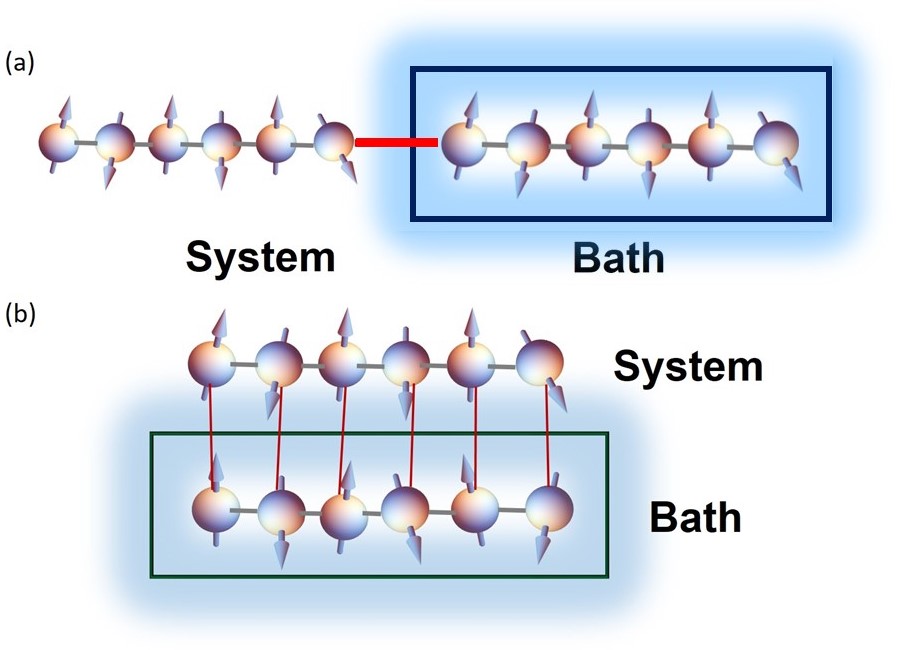}
	\caption{A schematic depiction of a disordered spin chain interacting with a bath. We consider two types of system-bath configurations: (a) the system is connected to the bath with a single connection and (b) the system is attached to the bath in a ladder configuration.}
	\label{fig:cartoon1}
\end{figure}

We numerically investigate these questions by studying the dynamics of a prototypical system exhibiting MBL - the disordered $s=1/2$ Heisenberg chain \cite{pal2010many, luitz2015many} - when specific initial states are coupled together under different configurations (either in a \textit{junction} or \textit{ladder} configuration - see Fig.~(\ref{fig:cartoon1})). By investigating these configurations for a range of parameters and system sizes, we wish to understand how the strength and geometry of couplings to a small bath affects the localisation properties of both the system and the bath.
Allowing two chains of differing disorder to interact with different interaction strengths, we study their dynamics and steady-state localisation properties, using the staggered magnetisation as a diagnostic of localization \cite{smith2016many, hauke2015many,wu2016understanding}.


Our numerical results indicate that the resulting localization properties arise from a complex interplay of disorder strength, system-bath geometry, coupling, and the initial states. In the junction configuration, both the system and the environment tend to retain their localisation properties, independent of the system-environment interaction strength.
For the ladder configuration, we observe a similar conclusion for system-environment interaction strengths that are comparable or smaller than the ladder's intra-chain interaction strength - there is no domination of localization or ergodicity, only a weakening of both characteristics. Numerics on larger system sizes support this conclusion in both configurations. On the other hand, for large interaction strengths, the system and environment is characterised by the dimerisation of each 2-qubit ladder rung. The eventual dynamics and steady-state behaviour of the dimerised chain is then dependent on initial conditions, which can tend towards localisation or thermalization, which we illustrate with analyses on the dynamics of individual spins.
Our results supplement existing studies on MBL systems coupled to small baths \cite{hyatt2017many, nandkishore2015many, wybo2020entanglement,rubioabadal2019many,luitz2017small,goihl2019exploration,kelly2020exploring,morningstar2021avalanches}, and generalizes them by exploring the effects of different geometries, interaction strengths, environment disorders, and nonequilibrium initial states. Using an experimentally accessible diagnostic of localisation, we hope to shed light on the interplay between localization and ergodicity, and the potential for localized systems to encode information.

We structure the article as follows : in Section \ref{sec:sysenv}, we describe the model and the two interaction configurations considered. In Section \ref{sec:signatures}, we define and motivate the use of staggered magnetisation as a dynamical signature to detect localization. Next, we present our main numerical results in Section \ref{sec:numerical}, along with additional numerics for different system sizes.
In Section \ref{info_retention}, we discuss the applications of our results in quantum memory. 
Finally in Section \ref{sec:conclude}, we conclude and reiterate our main results. 

\section{System-environment model}\label{sec:sysenv}
The total Hamiltonian of the system and its environment can be written as:
\begin{eqnarray}  \label{eq:total_h}
H_{tot} &=& H_s + H_e + H_{int},
\end{eqnarray}
where $H_s$, $H_e$ and $H_{int}$ are the system, environment and interaction Hamiltonians, respectively. 
In the following analyses, we take the system and environment to be 1D isotropic Heisenberg spin-1/2 chains of $l=6$ spins (forming a total of $L=2l=12$ spins) with disordered transverse magnetic field and open boundary conditions:
\begin{eqnarray}  
H_s &=& J \sum_{i=1}^{l-1} \vec{S}_i \cdot \vec{S}_{i+1} + \sum_{i=1}^{l} h_i^{(s)} S_i^z, \label{eq : sys hamiltonian} \\  
H_e &= & J \sum_{i=l+1}^{2l-1} \vec{S}_i \cdot \vec{S}_{i+1} + \sum_{i=l+1}^{2l} h_i^{(e)} S_i^z,  \label{eq : env hamiltonian}
\end{eqnarray} 
where $\vec{S}_i = (S_i^x, S_i^y, S_i^z)$ is the vector of local spin operators at site $i$, with $i \in [1, l]$ denoting the 6 system spins and $i \in [l+1, 2l]$ denoting the 6 environment spins. $J$ is the intra-chain interaction strength while $h_i^{(s)}$ ($h_i^{(e)}$) is the disorder parameter of the system (environment), which is a random real number uniformly distributed in the interval $[-W_s, W_s]$ ($[-W_e, W_e]$). Taken independently, the Hamiltonians Eq.~(\ref{eq : sys hamiltonian}) and Eq.~(\ref{eq : env hamiltonian}) have been extensively studied, with an ergodic-MBL transition known to occur at $W \approx 3.5J$ \cite{pal2010many, luitz2015many}.

The system and its environment are then allowed to interact via spin-spin interactions in two configurations:  
\begin{enumerate}
    \item \textbf{Junction configuration:}
    \begin{equation} \label{eq : junc_int}
        H_{int}^{(junc)} = J_{int} \vec{S}_l \cdot \vec{S}_{l+1}
    \end{equation}
    \item \textbf{Ladder configuration:}
    \begin{equation} \label{eq : ladder_int}
        H_{int}^{(ladder)} = J_{int} \sum_{i=1}^{l} \vec{S}_i \cdot \vec{S}_{i+l}
    \end{equation}
\end{enumerate} where $J_{int}$ represents the strength of the system-environment coupling. 
Fig.~(\ref{fig:cartoon1}) schematically illustrates these two system-bath configurations.
Throughout the article, we set $J=1$. We will also set the disorder of the system to be $W_{s} = 9J$ so that the system lies well within in the MBL regime, while allowing the environment disorder $W_{e}$ to vary from $W_{e} = 0$ (representing an ergodic phase) to $W_{e} = W_{s}$ (representing a strongly localized phase).

\section{Dynamical signatures of localisation}\label{sec:signatures}
An important signature of MBL systems is the existence of a set of local conserved operators, referred to as quasilocal integrals of motion (LIOMs). This integrability constrains the system's dynamics, leading to atypical dynamical properties such as a logarithmic growth of entanglement and boundary-law scaling of entanglement entropy \cite{bauer2013area, serbyn2013local, huse2014phenomenology}. The dynamical signature that we focus on is the equilibration of local observables to nonthermal values under a quantum quench \cite{pal2010many, serbyn2014quantum}, in particular that of the local physical spin operators $S_i^z$. The LIOM theory predicts that at long times, $\langle S_i^z \rangle$ will equilibrate to values that carry information about the initial state. This is in contrast to thermalizing systems which retain no long term memory, evolving into a temperature dependent equilibrium state with no memory of the initial states. 

As we will subsequently restrict our attention to initial N\'eel states of the form $\ket{\uparrow \downarrow \uparrow \downarrow ...}$ (chosen as an instance of a highly non-thermal initial state), to measure the collective effect of the equilibration of $\langle S_i^z \rangle$ across the entire chain, it is instructive to measure the (normalised) staggered magnetisation, defined as:
\begin{equation} \label{eq : stag_mag_expt}
    M(t, [a, b]) = \frac{1}{|b-a|} \sum_{i=a}^{b} (-1)^i \bra{\psi(t)} S_i^z \ket{\psi(t)},
\end{equation}
where $1/|b-a|$ is a normalization factor, and $[a, b]$ is the portion of subsystem of interest. We further denote $M_{sys} \equiv M([1, l])$ and $M_{env} \equiv M([l+1, 2l])$ the staggered magnetisation of the system and environment chains respectively. As a probe of localization, the staggered magnetisation has the advantage of being experimentally accessible \cite{smith2016many}, and is equivalent to the particle imbalance probed in cold-atom setups \cite{schreiber2015observation}.

We will be interested in the disorder-averaged quasi-steady state behaviour of $M$ at late times, i.e. the behavior of:
\begin{equation} \label{eq : ss_stagmag}
    \overline{M}([a, b]) = \frac{1}{\Delta t_{SS}} \int_{t_{SS}} M(t, [a, b]) dt
\end{equation}
when it is averaged over numerous disorder realizations, with an appropriately chosen steady-state window $t_{SS}$ that does not contain transient behaviour. The bar denotes averaging over the steady-state window, and we will always consider disorder-averaged quantities (the notation of which we suppress). We thus expect that if an initial highly non-thermal state of the form $\ket{\psi(0)} = \ket{\uparrow \downarrow \uparrow \downarrow ...}$ is subjected to a quench, $\overline{M}$ should remain close to its initial value of $\overline{M} = 1$ if the system is fully localized, while decaying to $\overline{M} = 0$ in the ergodic case. 

The staggered magnetisation can also be rescaled to define the Hamming distance:
\begin{equation}
    D(t) = \frac{1}{2}(1-M(t)),
\end{equation}
which is 0.5 for a thermal state and 0 for a fully localized N\'eel state. Also a quantifier of localization, it has been studied in MBL models \cite{hauke2015many,wu2016understanding}, including an experimental demonstration of MBL in a disordered long-range Ising model with ultracold ions \cite{smith2016many}. For initial states that are not in the N\'eel form, a more general measure is desirable, and in Appendix \ref{sec:fidelity} we study the fidelity, which is a quantifier of localization that does not depend upon the initial state being in the N\'eel form. Alternatively, one may use state-independent generalizations of the staggered magnetization that reduces to Eq.~(\ref{eq : stag_mag_expt}) for the N\'eel state as in Ref. \cite{guo2021stark,morong2021observation}.

To compute the long-time averages of observables, in lieu of an evolution over a finite duration and averaging over a window at late times as we have done, we note that one can in principle obtain the infinite-time average with the diagonal ensemble \cite{rigol2008thermalization} (This is explored in Appendix.~(\ref{sec:diagonal_ensemble})). However, the latter approach requires an exact diagonalization of the Hamiltonian, which we observed to be more computationally demanding than an evolution in a closed system for finite time. Since we observe good agreement in values and qualitative behaviour from both approaches (See Fig.~(\ref{fig:junc_dynamics})), subsequent analyses will be done with the former approach.

\section{Numerical results}\label{sec:numerical}
In this section, we present numerical results on the dynamics under the junction and ladder configurations, and investigate features that arise.
To study the steady-state localization properties of the system, we initialize  the total system in a highly non-thermal initial product state $\ket{\psi(0)} = \ket{\psi^{s}(0)} \otimes \ket{\psi^{e}(0)} $, allow it to evolve under $H$, and study the subsequent dynamics of $\ket{\psi(t)}$. Each individual chain starts as a N\'eel state, and we denote $\ket{\psi_{even}} \equiv \ket{ \downarrow \uparrow \downarrow \uparrow \downarrow \uparrow}$ and $\ket{\psi_{odd}} \equiv \ket{\uparrow \downarrow \uparrow \downarrow \uparrow \downarrow}$ to distinguish between N\'eel states with positive spins at even-numbered sites and those with positive spins at odd-numbered sites respectively. This distinction is important for the localization dynamics of the ladder configuration, as shown in Section \ref{Ladder}.
To quantify the extent of localization or the preservation of the initial state, we study the staggered magnetisation $\overline{M}$ as defined in Eq.~(\ref{eq : ss_stagmag}). It measures the deviation of $\ket{\psi(t)}$ from the initial N\'eel state. 

\subsection{Junction} \label{Junction}
\begin{figure}
     \subfloat[]{\label{junction_dynamics}%
         \includegraphics[width=.24\textwidth]{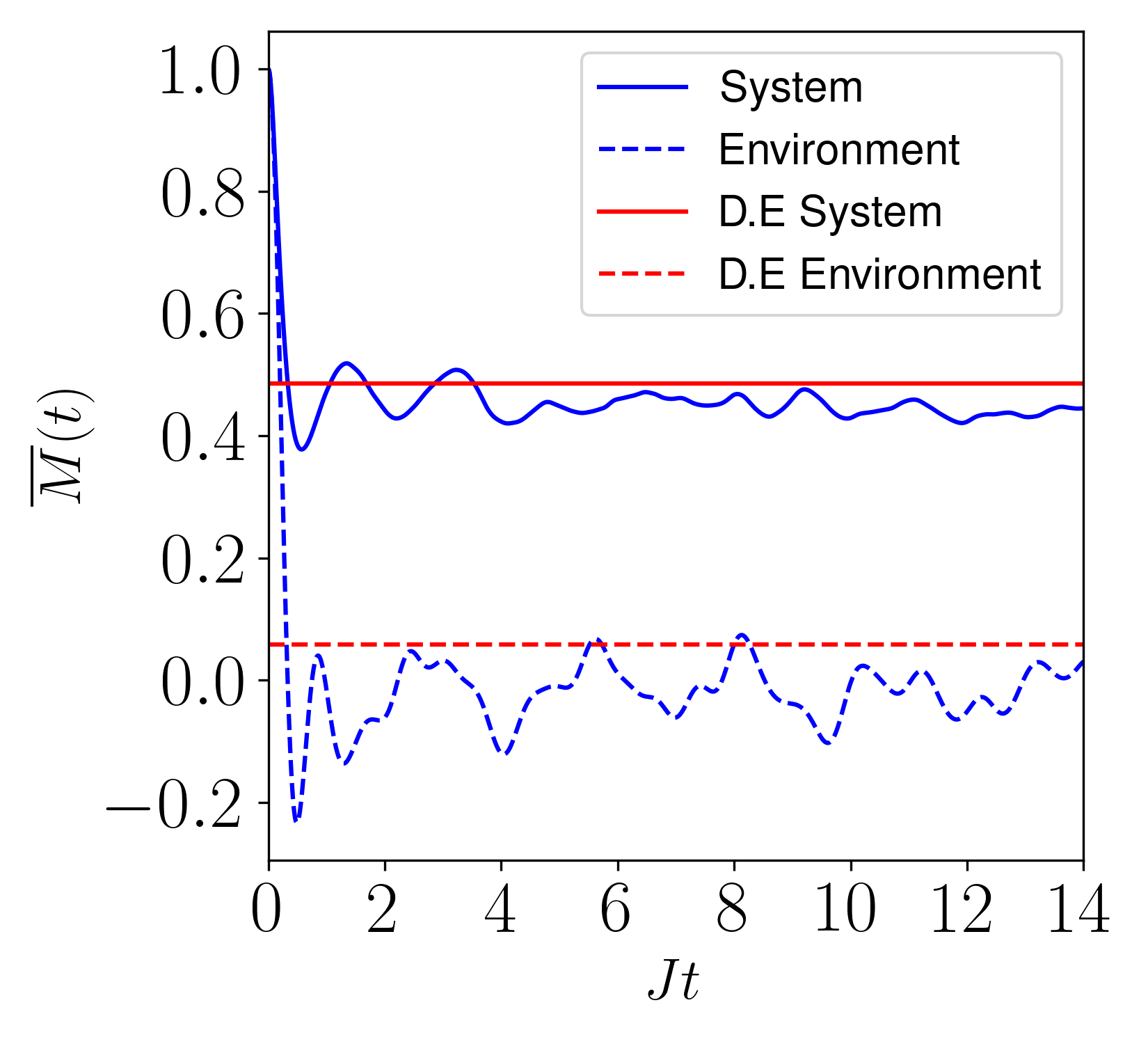}
         }
     \subfloat[]{\label{junction_spins}%
         \includegraphics[width=.25\textwidth]{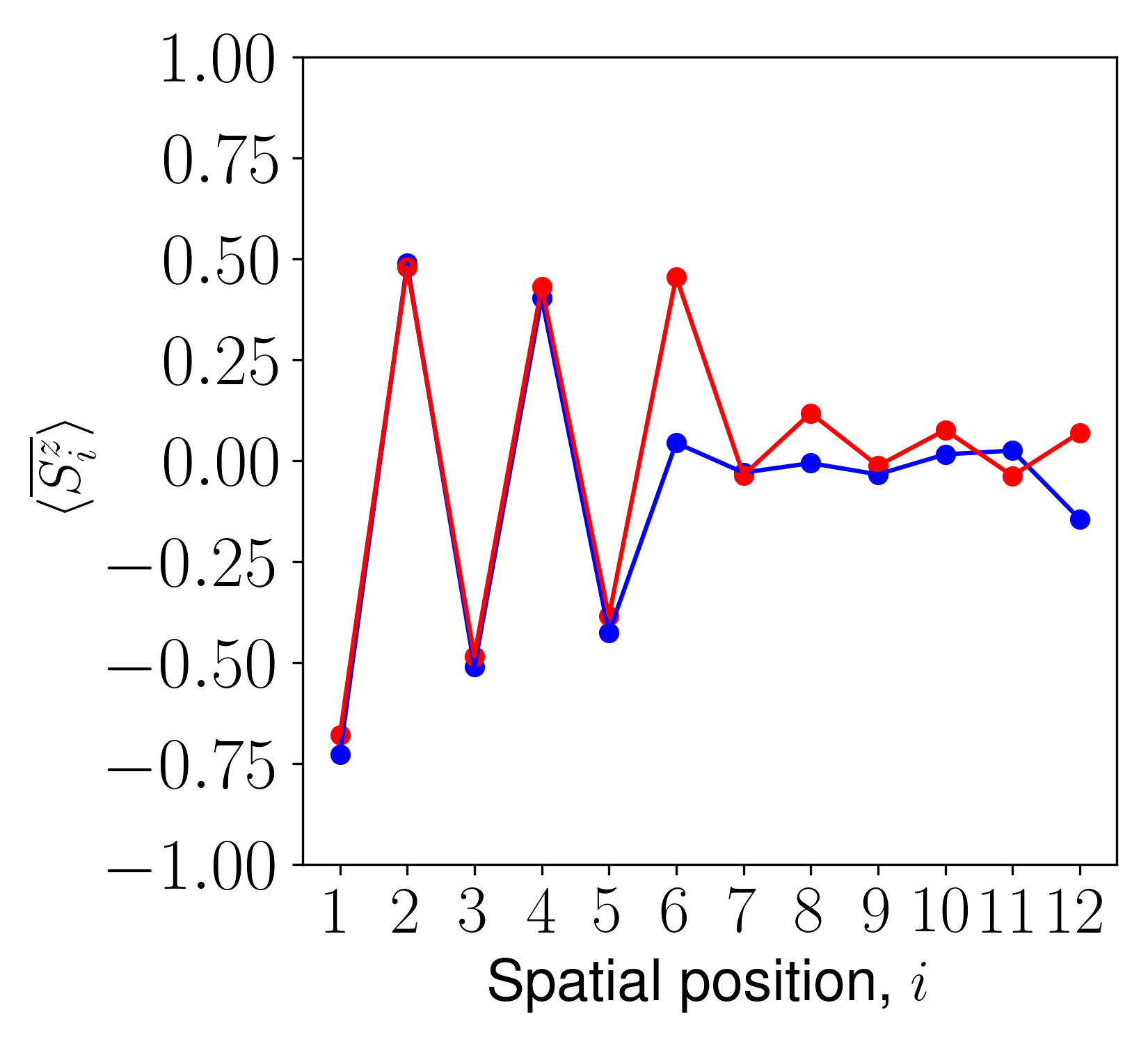}
         }
\caption{ (a) Dynamics of staggered magnetisation, $M(t)$, and (b) steady state spatial distribution of local spin expectation values for parameters $J_{int} = J = 1$, $W_{env} = 0$, and $W_{sys} = 9$, averaged over 100 disorder realisations. The steady-state window $\Delta t_{ss}= 2.4$ corresponding to $t \in [9.6,12]$ is used to compute the steady-state spin distribution of (b). Error bars are too small to be displayed. The red line in Figs (a) and (b) indicates staggered magnetisation and local spin values obtained with the diagonal ensemble respectively (See Appendix.~(\ref{sec:diagonal_ensemble})).}
\label{fig:junc_dynamics}
\end{figure}

\begin{figure}
     \centering
     \includegraphics[width=\linewidth]{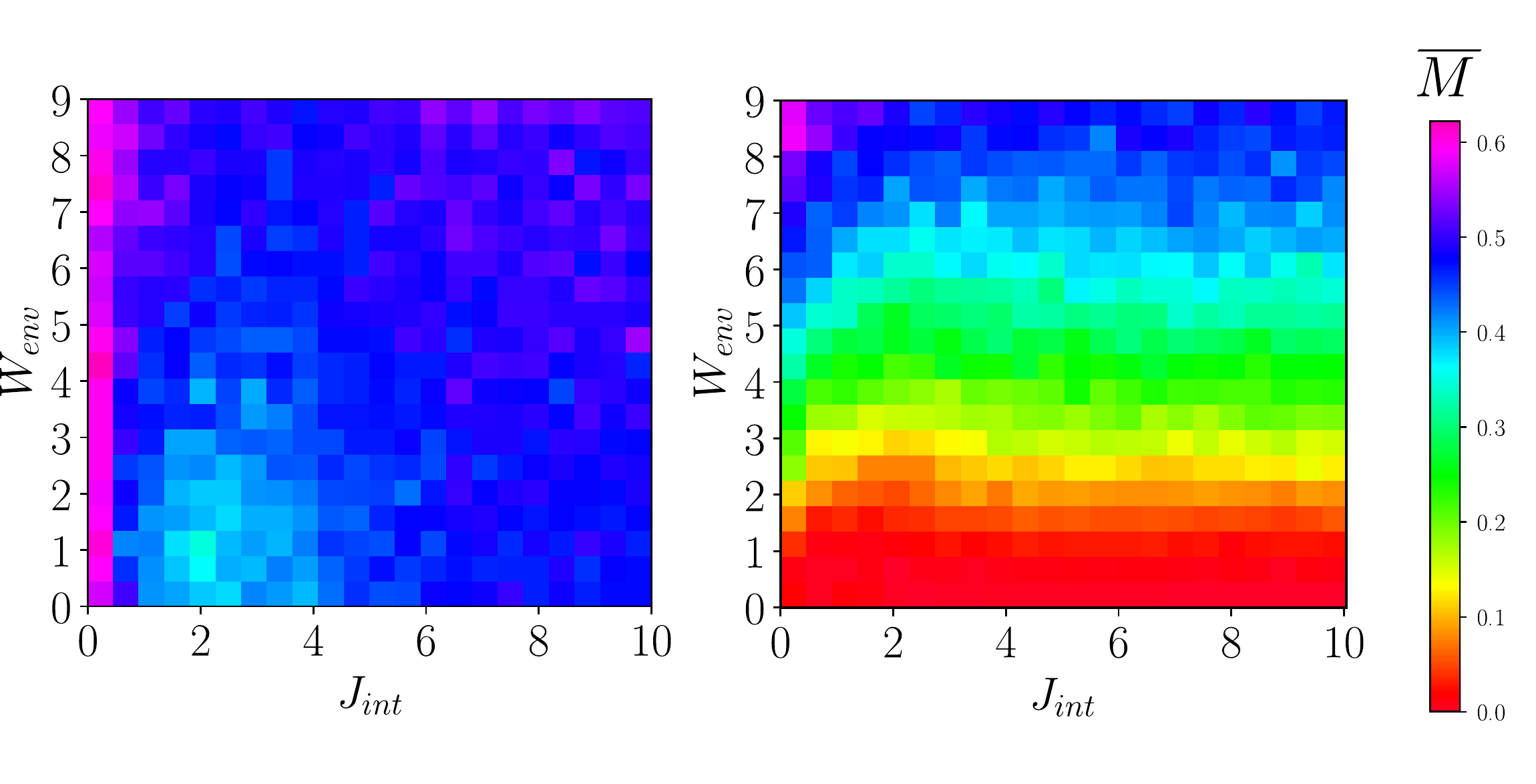}
\caption{Steady-state staggered magnetisation of the system, $\overline{M}_{sys}$ (left), and environment, $\overline{M}_{env}$ (right), for the junction configuration. The tuples $(J_{int}, W_{env})$ form a grid separated by 0.5 intervals. Each point is averaged over 200 disorder realisations, evolved for $Jt = 50$.}
\label{fig:junc}
\end{figure}

We first consider the junction configuration and the initial state $\ket{\psi(0)} = \ket{\psi^{s}_{even}} \otimes \ket{\psi^{e}_{even}}$. In this case, the ergodic and localized 1D subsystems interact through a single contact point that does not scale with system size. An understanding of the eventual localization properties of this configuration is crucial as it relates to the effects of Griffiths regions and their impact on the stability of MBL in 1D, which is an ongoing subject of study \cite{de2017stability,luitz2017small,morningstar2021avalanches}.

To illustrate the quenched dynamics and equilibration at long times of the staggered magnetization, we consider a strongly localised system ($W_{sys} = 9$) coupled to a strongly ergodic environment ($W_{env} = 0$) with inter-chain interaction strength $J_{int} = J$. Fig.~(\ref{junction_dynamics}) shows the dynamics of $M_{sys}$ and $M_{env}$. Starting from the initial state with $M_{sys}(t=0) = M_{env}(t=0) = 1$, both curves display a rapid decrease in $M$ at short times, before fluctuating about different equilibrium values ($M_{sys} \approx 0.4$ and $M_{env} \approx 0$) after a transient period. Fig.~(\ref{junction_spins}) further shows the equilibrated spatial distribution of local spin expectation values $\langle S_i^z \rangle$, averaged over the steady-state window $t \in [9.6, 12]$, appropriately chosen beyond the transient period.

The large value of $M_{sys} \approx 0.4$ reflects the preservation of information in the localized system half-chain despite its coupling to an ergodic environment. On the other hand $M_{env} \approx 0$ signals information loss in the other ergodic environment half-chain. This can also be seen from the spin distribution in Fig.~(\ref{junction_spins}) - the system retains its alternating local spin expectation values $\langle S_i^z \rangle$, while the environment retains no such feature. 

We also note that the spin belonging to the system closest to the boundary ($i=6$) has equilibrated to a value close to zero, signifying the penetration of ergodicity into the localized system. While spins close to the boundary contribute to the value of the equilibrated staggered magnetisation, this contribution is expected to decrease as the length of the chain is increased.

Under these parameters, we thus observe that both the system and environment retain their localization properties despite the coupling. That is, they evolve as independent half-chains, even with contrasting localisation properties, as the only interaction between the two chains occur at the boundary. 

Indeed, we find that this conclusion can be further extended to a range of values for $J_{int}$ and $W_{env}$, as we summarize in Fig.~(\ref{fig:junc}), which shows the values of $\overline{M}_{sys}$ (left) and $\overline{M}_{env}$ (right) for $J_{int} \in [0, 10]$ and $W_{env} \in [0, 9]$. We note the large values of $\overline{M}_{sys}$, regardless of $J_{int}$ and $W_{env}$ - the system retains its localisation (in particular, even when the system is interacting strongly with a highly ergodic bath). 

Another interesting observation is that the system-environment coupling shifts the apparent ergodic-MBL transition point to larger disorder values.  This is apparent by comparing the values of $\overline{M}_{env}$ at $J_{int} = 0$ with finite $J_{int}$ in Fig.~(\ref{fig:junc}).

We expect the above observations to persist for periodic boundary conditions and increasing system sizes, as the number of interactions do not scale with chain length. This is confirmed in Subsection.~(\ref{subsec:scaling}), where additional numerical simulations on different total system sizes display the retention of localization properties in both the system and environment. With increasing system size, the contribution of the system-environment coupling to $\overline{M}$ will also tend to zero, visible in Fig.~(\ref{junction_spins}) at sites 6 and 7, and in Fig.~(\ref{fig:scaling_L}).

\subsection{Ladder} \label{Ladder}
Next, we consider a more physically realistic situation, in which the system-bath interactions scale with system sizes. One possible configuration is the ladder configuration described by $H = H_s + H_e + H^{(ladder)}$ from Eqs.~(\ref{eq : sys hamiltonian}), (\ref{eq : env hamiltonian}) and (\ref{eq : ladder_int}) in which every spin in the system interacts with an environment spin with interaction strength $J_{int}$ (illustrated in Fig.~(\ref{fig:cartoon1}b)). 

It is known that the physical properties of this configuration depends on the ratio $\gamma \equiv J/J_{int}$~\cite{bouillot2011statics}. In the limit where $J_{int} = 0$, the two chains decouple and evolve independently of one another. A small but non-zero $J_{int}$ then acts as a weak perturbation to each chain. On the other hand, in the limit where $J = 0$ and so $\gamma = 0$, each ladder rung decouples from one another and evolves independently under the local rung Hamiltonian:
\begin{equation} \label{eq : ladder rung}
    H_i = J_{int} \vec{S}_i \cdot \vec{S}_{i+l}.
\end{equation}
The eigenstates of (\ref{eq : ladder rung}) are the triplets $\ket{\uparrow \uparrow}$, $\ket{\downarrow \downarrow}$, $\frac{1}{\sqrt{2}}( \ket{\uparrow \downarrow} + \ket{\downarrow \uparrow})$ with total spin 1, and the singlet state $\frac{1}{\sqrt{2}}( \ket{\uparrow \downarrow} - \ket{\downarrow \uparrow})$ with total spin 0 - each rung is effectively a site that can exist as a spin-0 or spin-1 particle (or a superposition of both).

In the following, we again follow a similar analysis as the previous section by studying the quenched dynamics of an initial state $\ket{\psi(0)}$ evolving under the Hamiltonian $H = H_s + H_e + H^{(ladder)}$. Fixing $J=1$, we start in the limit $J_{int}=0$ where both chains are decoupled and vary the ratio $\gamma = J/J_{int}$ by increasing $J_{int}$. 

\begin{figure}
     \centering
         \includegraphics[width=\linewidth]{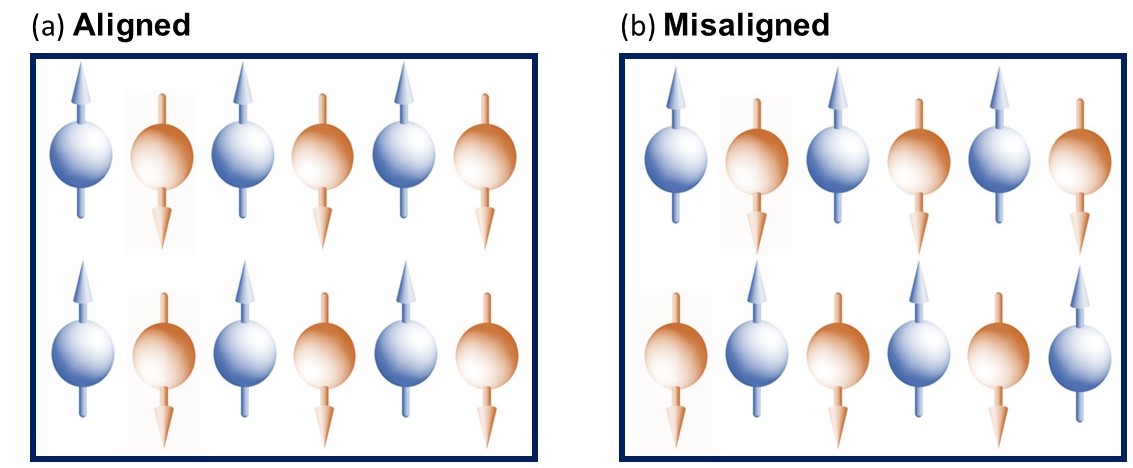}
\caption{Illustration of different initial states for the ladder configuration: (a) Aligned state, $\ket{\psi_A}$ and (b) Misaligned state, $\ket{\psi_B}$}
\label{fig:initial_states}
\end{figure}

\begin{figure}
     \centering
         \includegraphics[width=\linewidth]{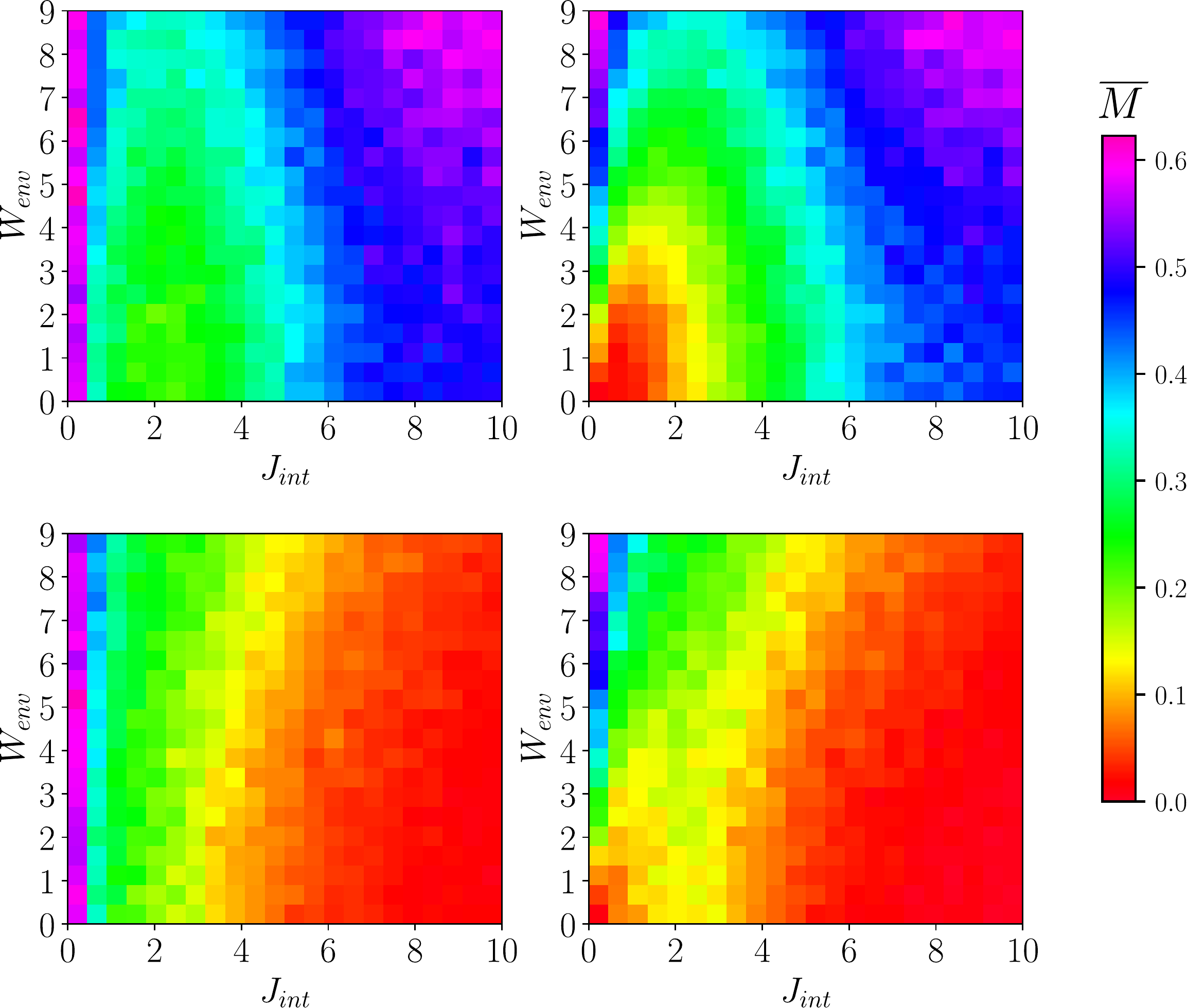}
\caption{Steady-state staggered magnetisation for the system, $\bar{M}_{sys}$ (left), and environment, $\bar{M}_{env}$ (right), for the ladder configuration. The upper two plots were obtained with the aligned initial state $\ket{\psi_A} \equiv \ket{\psi_{even}} \otimes \ket{\psi_{even}}$, while the bottom two plots were obtained with the misaligned initial state $\ket{\psi_M} \equiv \ket{\psi_{even}} \otimes \ket{\psi_{odd}}$. The tuples $(J_{int}, W_{env})$ form a grid separated by 0.5 intervals. Each point is averaged over 200 disorder realisations, evolved for $Jt = 50$. }
\label{fig:ladder}
\end{figure}

The subsequent results indicate that the steady-state localization properties of the system will persist for small to intermediate inter-chain coupling strength $J_{int}$, but at large $J_{int}$ different dynamics emerge, which depends strongly on the initial state of both the system and the environment. We will consider two different initial states: we call $\ket{\psi_A} \equiv \ket{\psi_{even}} \otimes \ket{\psi_{even}}$ the aligned state, and $\ket{\psi_M} \equiv \ket{\psi_{even}} \otimes \ket{\psi_{odd}}$ the misaligned state (See Fig.~(\ref{fig:initial_states})). Importantly, the two initial states have different energy densities due to the interaction terms $H_{int}^{(ladder)}$.  This term is positive for the aligned case and negative for the misaligned case. In the regime where $J_{int}$ is large, its contribution to the energy density becomes significant, resulting in distinct dynamics for the two different initial states.

Fig.~(\ref{fig:ladder}) shows the values of $\overline{M}_{sys}$ (Left) and $\overline{M}_{env}$ (Right) for $J_{int} \in [0, 10]$ and $W_{env} \in [0, 9]$, for both choices of initial states. In the following, we analyse our results in the weak, intermediate and strong system-environment interaction regimes independently. 

\subsubsection{Weak interaction regime : $ J_{int} \ll J$} 
For $J_{int}$ much smaller compared to $J$, the system and environment are reduced to independently evolving chains that weakly perturb one another. This is consistent with our expectations, illustrated by the independence between $M_{sys}$ and $W_{env}$ in Figs.~(\ref{fig:ladder}) at small values of $J_{int}.$ For $J_{int}=0$, when $W_{env}$ is increased, the environment transitions from being ergodic to localised, while the localised system remains localised, with $M_{sys}$ unchanged.

\subsubsection{Intermediate regime : $ J_{int} \approx J$} \label{subsec:intermediate}
In this regime, the ladder configuration represents an intermediate configuration between 1D and 2D. In the case of $J=J_{int}$, this has been studied in the context of MBL to exhibit localisation beyond a critical disorder strength across the entire ladder \cite{baygan2015many}. Since our focus is on the coupling between two systems of different localisation properties, the above situation is effectively a special case of our results when $W_{env} = W_{sys}$.

For $J \approx J_{int}$, while $\overline{M}_{env}$ responds to changes in $W_{env}$, the near constant value of $\overline{M}_{sys} \approx 0.3$ in Fig.~(\ref{fig:ladder}) indicates the preservation of localization in the system, regardless of the environment disorder. The system and environment thus evolve independently from one another, illustrated by the strong independence between $\overline{M}_{sys}$ and $\overline{M}_{env}$. 
In particular, this observation is consistent with the fact that the critical disorder strength for a ladder with $J_{int} = J$ has been determined to be $W = 8.5 \pm 0.5 J$ by analyses on the scaling of entanglement entropy and spectral statistics \cite{baygan2015many}. This parameter choice corresponds to points close to the upper-left corner of the plots of Fig.~(\ref{fig:ladder}), i.e the points $(J_{int}, W_{env}) = (1, 8.5)$.  

We note the drop of $\overline{M}_{sys}$ when the system is coupled to the environment at $J \approx J_{int}$, which is independent of the environment disorder $W_{env}$. Consistent with the ladder's larger critical disorder strength of $W = 8.5 \pm 0.5 J$, this is because the two chains effectively form a quasi 2D configuration, which localizes at a larger disorder.

We conclude that in this regime, the initially localized system and environment both retain their localization properties upon coupling, albeit to a lesser extent than if they were connected via the junction configuration. This degradation can be attributed to the transition from a 1D system to a quasi 2D one, which requires larger disorder to be localized. We perform additional simulations in this regime for different total system sizes in Subsection.~(\ref{subsec:scaling}), and we find that the above conclusion continues to hold.

\subsubsection{Strong interaction regime : $ J_{int} \gg J$} \label{strong_int_reg}
For large values of $J_{int}$, Fig.~(\ref{fig:ladder}) indicates that  $\overline{M}_{sys}$ and $\overline{M}_{env}$ become strongly correlated, with $\overline{M}_{sys} \approx \overline{M}_{env}$. The boundary at which this occurs is seen  from the figures at values of $J_{int} \in [4, 6]$, depending on $W_{env}$. The values of $\overline{M}$ also depends now on the initial state, with $\overline{M} \approx 0.55$ for the aligned initial state (blue regions in top plots) and $\overline{M} \approx 0$ for the misaligned initial state (red regions in bottom plots). 

The values of $\overline{M}_{sys}$ and $\overline{M}_{env}$ in each case can be explained by  the large interaction strength $J_{int}$ between qubits in the top and bottom rung reducing the ladder into a series of dimers that interact with their neighbors. That is, in the timescale $Jt$, the spin-1/2 ladder is reduced to a single spin chain consisting of particles that have both spin-0 and spin-1 degrees of freedom. The quenched dynamics of the ladder thus depends strongly on the rungs' initial configuration.

 \begin{figure} 
    \centering
    \includegraphics[width=.48\textwidth]{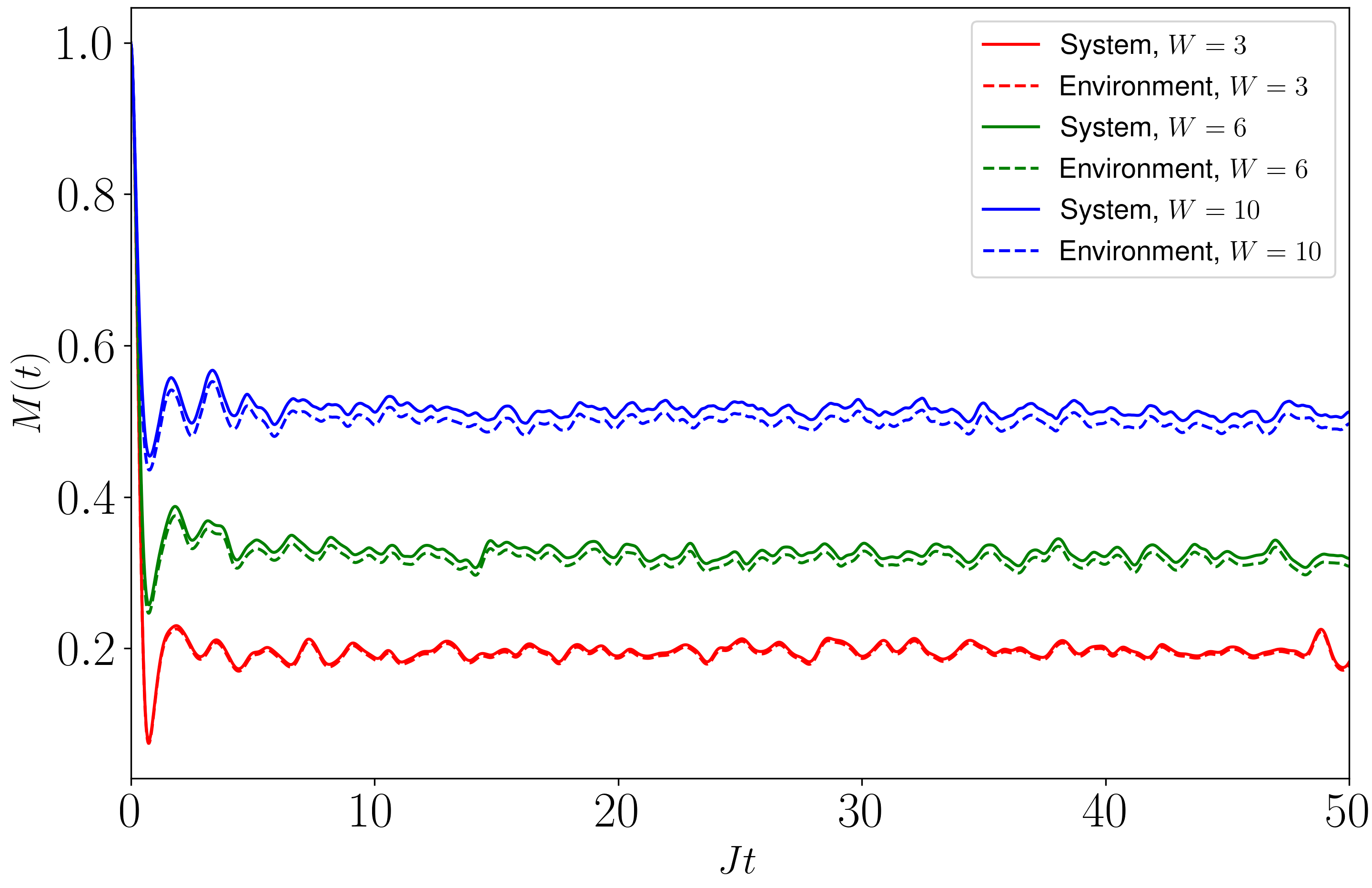}
\caption{
Dynamics of staggered magnetization for the aligned initial state in the ladder configuration, for different disorder $W$ and interaction strength $J_{int} = 20$. Solid line indicates system staggered magnetisation $M_{sys}$ while dashed lines indicate environment staggered magnetisation $M_{env}$. When $J_{int}/J \gg 1$, the dynamics is qualitatively similar to that of a 1D chain, and $\langle S_i^z \rangle \approx \langle S_{i+l}^z \rangle$.}
\label{fig:ladder_dynamics_aligned}
\end{figure}

For an initial aligned state $\ket{\psi_A}$, each rung is either $\ket{\uparrow \uparrow}$ or $\ket{\downarrow \downarrow}$, which correspond to two of the triplet states. More generally, if the top and bottom spins of a rung are initially aligned in the same direction, it is an eigenstate of the $J_{int}$ term of the total Hamiltonian - the resulting two-qubit product state can always be written as a superposition of the triplets. An evolution generated by the same term will therefore only result in the multiplication of a global phase factor. As $J << J_{int}$, in the timescale $Jt$, the two spins of a rung are therefore strongly coupled, with a common direction that can effectively be described by a total spin vector. This total spin vector then interacts with its neighbors via the $J$ terms of the total Hamiltonian. In this case the ladder is reduced to a chain of interacting spin-1 particles with an effective external disorder that depends on both $W_{sys}$ and $W_{env}$, with the steady-state staggered magnetization indicating the strength of localisation arising from the effective disorder. This is the reason we observe apparent localization in the top two plots of Fig.~(\ref{fig:ladder}).

The dynamics of staggered magnetization for different values of $W_{env}$ shown in Fig.~(\ref{fig:ladder_dynamics_aligned}) illustrates this behaviour. For $J_{int} = 20$, the values of $M_{sys}$ and $M_{env}$ for each disorder strength $W$ are strongly correlated and close to one another, such that for larger $J_{int}$ we expect $M_{sys} \approx M_{env}$. Their dynamics and final steady-state values are thus qualitatively similar to that of a disordered spin-$1/2$ chain, where the staggered magnetisation scales with disorder strength, transitioning from thermal to localized. The large values of $M_{sys}$ and $M_{env}$ in Fig.~(\ref{fig:ladder}) also indicate that memory of the initial configuration is retained, encoded in the +1 and -1 total spin vectors of each dimer. We discuss and provide further numerical analysis on the reduction of the spin-1/2 ladder to a spin-1 chain in Appendix \ref{sec:stronglycoupled}.


 \begin{figure}
    \centering
    \subfloat[]{\includegraphics[width=.49\textwidth]{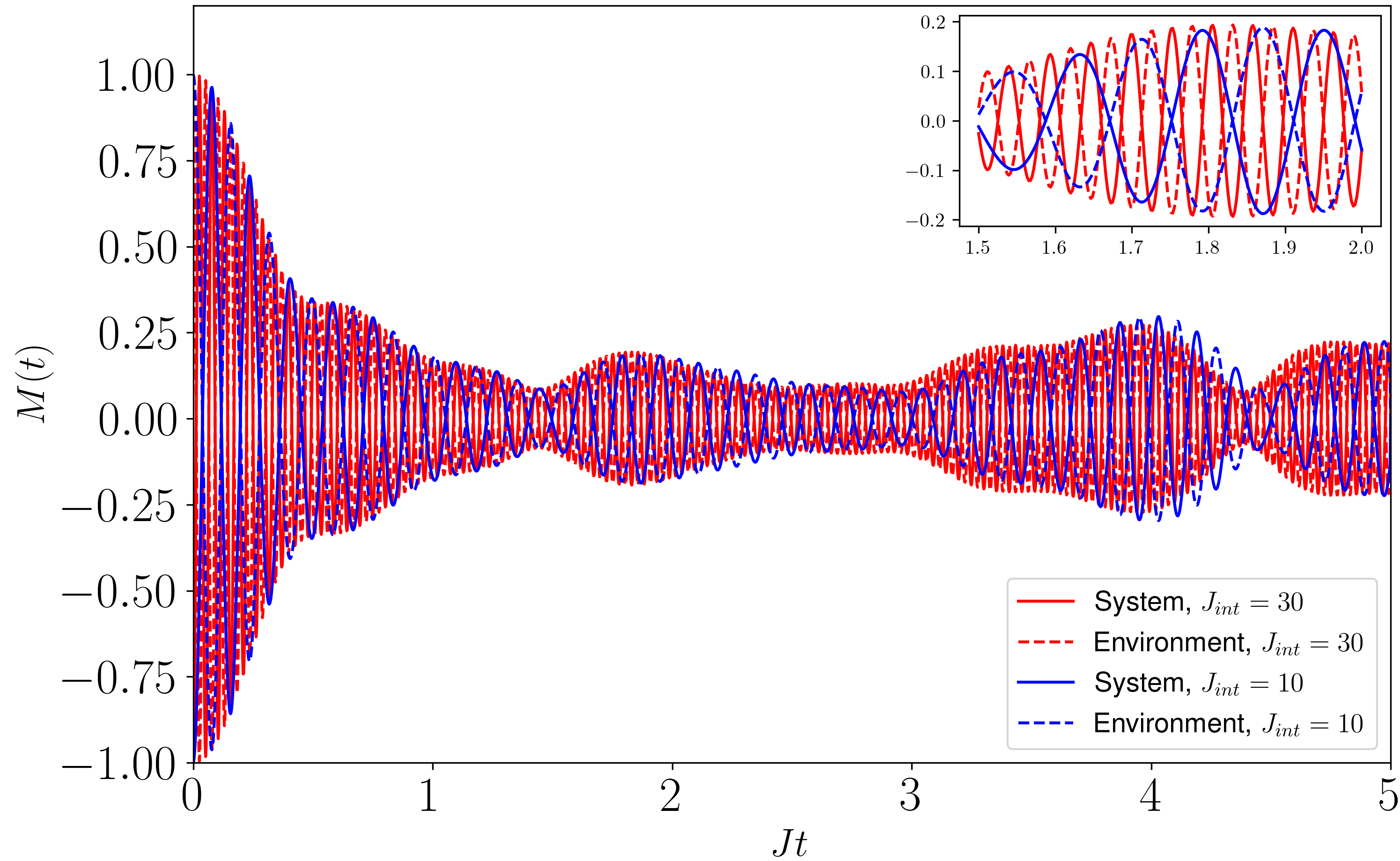}\label{fig:ladder_dynamics_misaligned} }
          
      \hfill
    \subfloat[]{\includegraphics[width=.49\textwidth]{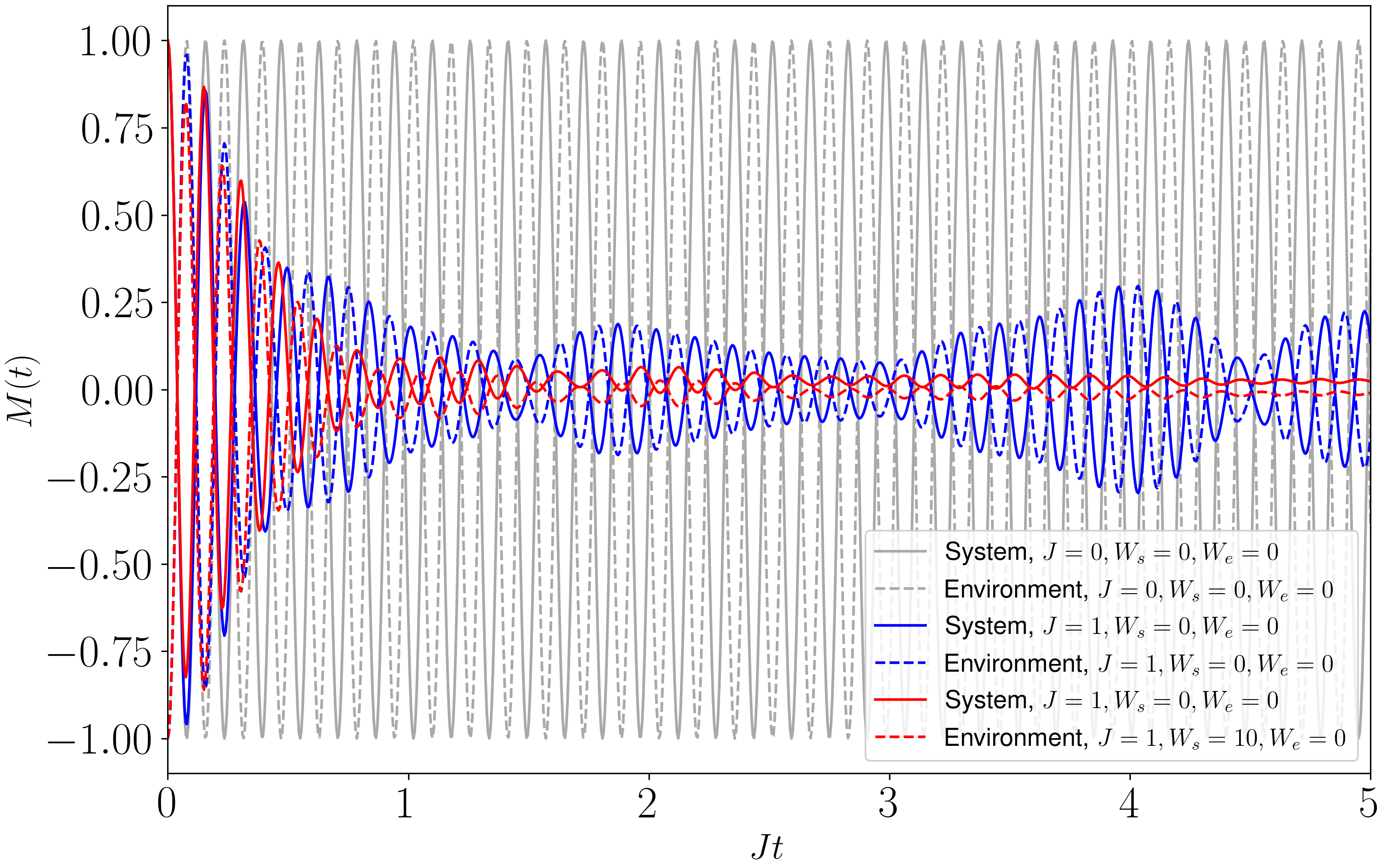}\label{fig:ladder_dynamics_misaligned_suppress}}
\caption{
Dynamics of staggered magnetization for the misaligned initial state in the ladder configuration, showing internal oscillations with frequencies proportional to $J_{int}$ (Top) and the suppression of amplitude due to disorder and neighboring interactions (Bottom).}
\end{figure}

For the misaligned initial state $\ket{\psi_M}$, each rung is either $\ket{\uparrow \downarrow}$ or $\ket{\downarrow \uparrow}$, which are superpositions of both singlet and triplet states. The dynamics in this case is separated into two timescales - the dynamics of the chain controlled by $J$, and the internal dynamics of each dimer controlled by $J_{int}$. This is shown in Fig.~(\ref{fig:ladder_dynamics_misaligned}), where the dynamics of staggered magnetisation consists of high frequency oscillations bounded by an envelope, with a time average of zero. The internal oscillations arise from the $J_{int}$ terms of the total Hamiltonian, which causes each dimer to oscillate between spin-0 and spin-1 modes with a frequency proportional to $J_{int}$ (See inset of Fig.~(\ref{fig:ladder_dynamics_misaligned})). On the other hand, the envelope originates from interactions between neighboring dimers due to the $J$ terms, and the presence of disorder. Both effects serve to introduce phase differences between the dimers, leading to the suppression of the amplitude of the envelope. Fig.~(\ref{fig:ladder_dynamics_misaligned_suppress}) shows this suppression effect as the neighboring interaction (controlled by $J$) and disorder (controlled by $W_{sys}$ or $W_{env}$) are successively turned on. Ultimately, the steady-state staggered magnetization averages to zero, leading to the red regions in Fig.~($\ref{fig:ladder}$) (Bottom) at large $J_{int}$. This situation is expected to occur whenever a two-qubit rung is in a superposition of spin-0 and spin-1 modes - the subsequent dynamics then consists of oscillations between the two spin degrees of freedom, resulting in apparent thermalization.

In a generic situation where the system and environment initial states are uncorrelated, we therefore expect apparent thermalization in this regime, leading to dynamics of the type shown in Fig.~(\ref{fig:ladder_dynamics_misaligned}). Only in the rare case where the spins are aligned in the same direction can dimerization occur, mapping the dynamics to that of a chain of spin 1 particles shown in Fig.~(\ref{fig:ladder_dynamics_aligned}). The onset of this regime is observed to be between $J_{int} \in [4,6]$. For larger system sizes, the same conclusions are expected to hold, due to the dimerization between top and bottom qubit pairs.

\subsection{Scaling for different total system sizes} \label{subsec:scaling}

\begin{figure}
    \centering
     \subfloat[]{\includegraphics[width=\linewidth]{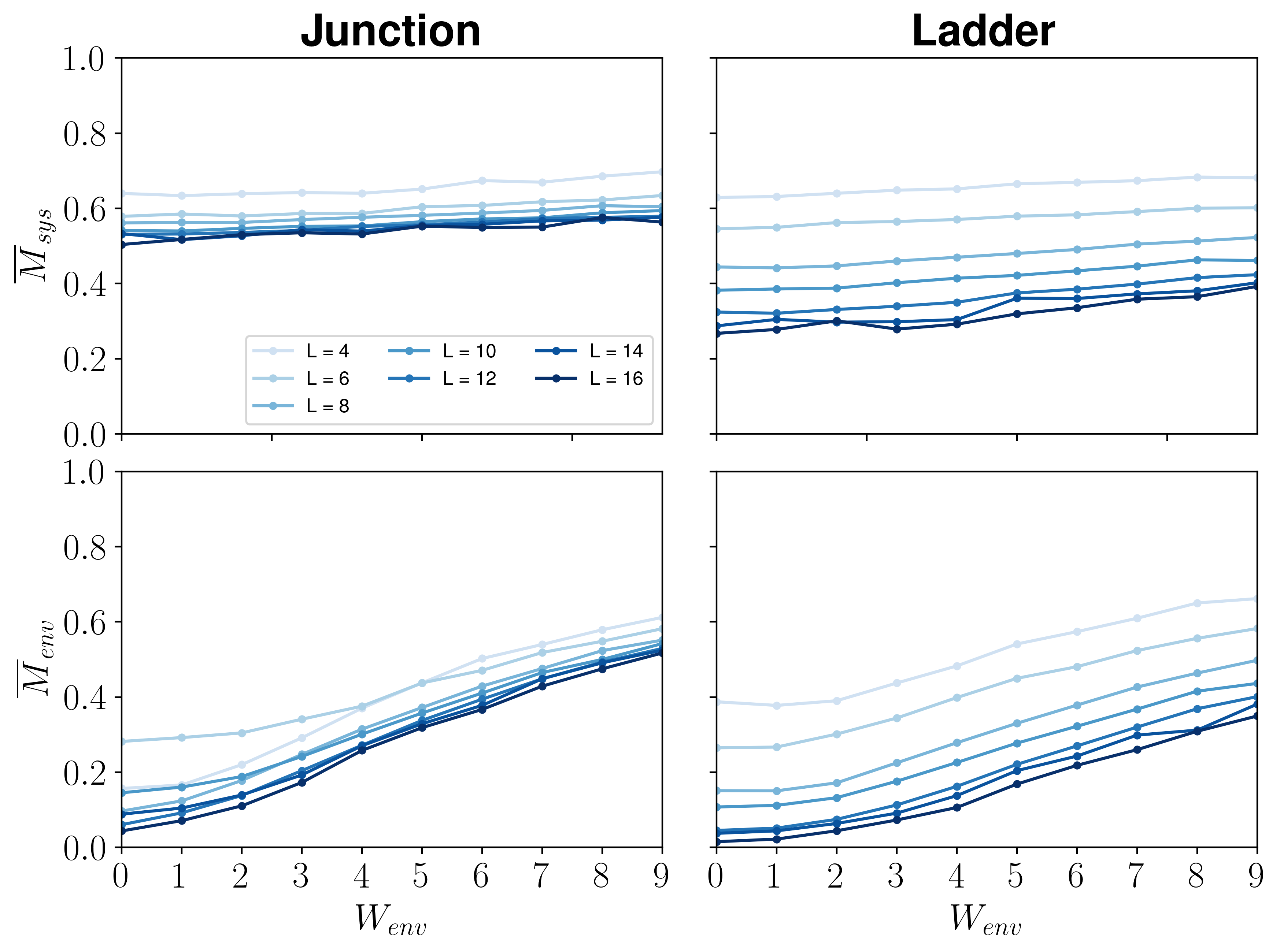}\label{fig:scaling_W}}
     \hfill
     \subfloat[]{\includegraphics[width=\linewidth]{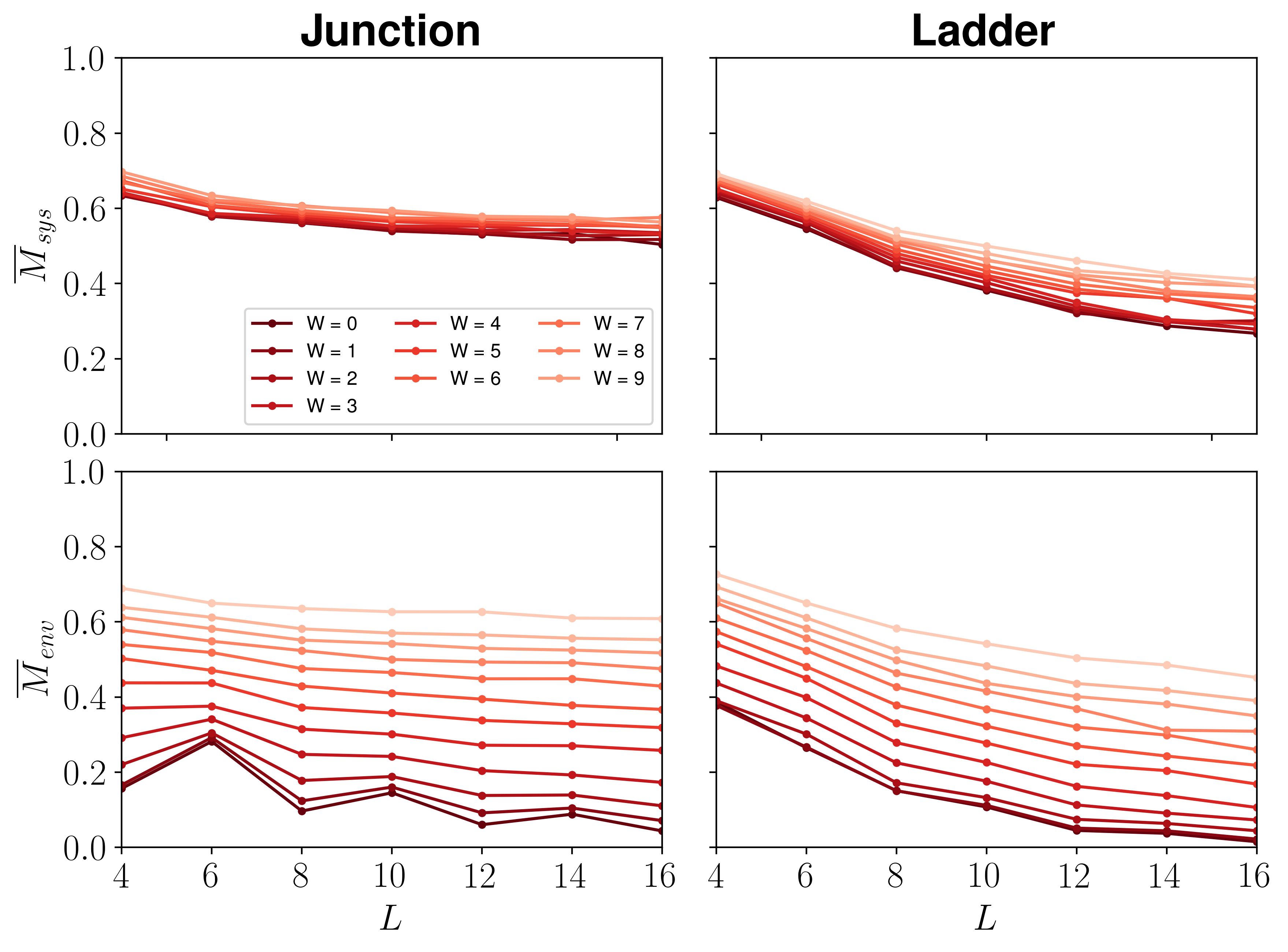}\label{fig:scaling_L}}
\caption{Steady-state staggered magnetisation $\overline{M}$ for different system sizes, for both the junction and ladder configurations in the intermediate regime $J_{int} = J = 1$. Top four curves of (a) in blue show response against $W_{env}$, while bottom four curves of (b) in red show response against $L$. Both sets of curves are plotted from the same dataset. We find similar qualitative results for the misaligned initial state, which we omit.}
\label{fig:scaling_fig} 
\end{figure}

In this section, we provide numerical evidence that similar conclusions hold in both the junction configuration of Subsection.~(\ref{Junction}) and the ladder configuration of Subsection.~(\ref{subsec:intermediate}) in the intermediate coupling regime for larger system sizes, and discuss the relation of this conclusion to other works.

Fig.~(\ref{fig:scaling_fig}) shows additional numerical results for different total system sizes of $L\in \{4, 6, 8, 10, 12, 14, 16 \}$. We consider both the junction and ladder configurations in the intermediate coupling regime $ J_{int} = J = 1$, with the system's disorder fixed at $W_{sys} = 9$, starting from the aligned initial state. (\ref{fig:scaling_W}) and (\ref{fig:scaling_L}) shows $\overline{M}_{sys}$ and $\overline{M}_{env}$ as functions of $W_{env}$ and $L$ respectively. 
For all system sizes, $\overline{M}_{sys}$ remains at nonzero values as $W_{env}$ is changed (top two blue plots), indicating that it remains localized regardless of the environment's localization properties. On the other hand, the environment shows a transition from ergodic to MBL as its disorder $W_{env}$ is changed (bottom two blue plots). The curves of $\overline{M}$ against $W_{env}$ also appear to be converging to a fixed function as $L$ increases, or equivalently that the gradients of the $\overline{M}$ against $L$ curves (red curves) are decreasing with increasing $L$. 
Taken together, these observations suggest that the main conclusion of the previous sections - that in the intermediate coupling regime, the system and environments both retain their localization properties - persists for larger system sizes.

It is instructive to compare these results with the related works of \cite{luitz2017small,goihl2019exploration}. In Ref.~\cite{luitz2017small}, one sees thermalization in an effective model with power law interaction (effectively a configuration that is intermediate between our junction and ladder configurations), while in \cite{goihl2019exploration} thermalization for the junction configuration with increasing system sizes was observed. Most relevant for us, in particular, is the observation in Ref.~\cite{goihl2019exploration} where one observes a slow transition to thermalization with increasing system sizes. Comparing with our junction results in Fig.~(\ref{fig:scaling_fig}), the distinction is apparent, where we instead observed clear signatures that a system coupled to a small bath does not thermalize ($\overline{M}_{sys}$ clearly remains at a large non-zero value for increasing system sizes (Red curves, top left plot))

The key distinction between our study and other studies such as \cite{luitz2017small,goihl2019exploration} is that most current work \cite{oganesyan2007localization,pal2010many,wybo2020entanglement,nico2020information} focuses on quantities such as entropies and the statistics of local observables which are taken over the entire eigenspectrum, while we study the steady-state properties of specific initial states. Our approach is also consistent with a large number of experimental investigations of MBL \cite{schreiber2015observation, smith2016many, choi2016exploring, bordia2017probing, xu2018emulating, guo2021observation}, since separable initial states such as the N\'eel state are easily prepared.

The difference in our conclusions indicates that while probes over the entire eigenspectrum may be more sensitive in detecting an ergodic-MBL transition, this general conclusion may not hold for initial states that are far from equilibrium such as the aligned and misaligned N\'eel states. A related issue is the presence of the MBL mobility edge \cite{luitz2015many,baygan2015many,li2015many,guo2021observation}, where states with different energy densities can give rise to different localization properties. This is illustrated in Fig.~(\ref{fig:ladder}), where dynamics under a MBL Hamiltonian with the same $W$ and $J_{int}$ results in significantly different magnetization values, depending on the energy density of the initial state.

This discussion highlights that conclusions gained from similar experimental and numerical results concerning the stability and scaling properties of the MBL phase should be interpreted with care, with the underlying dynamics and energy densities taken into account.

\section{Implications on information retention properties}
\label{info_retention}
In the actual implementation of many-body quantum systems for technological applications, unwanted dissipation and couplings to the environment cannot be completely eliminated. While a simple model treating the environment as an infinitely large system can be described by a quantum master equation, the neglect of backaction effects eventually drives the MBL system towards a thermal state logarithmically \cite{everest2017role, wybo2020entanglement}. As a first step in studying whether such a finite bath can lead to a different picture, it is instructive to consider both the system and the environment as a closed system.

Our numerical approach answers the question for the case when the system and environment belong to the same type and size. This is relevant for future implementations of many-body quantum systems; as isolation with the external environment with large number of degrees of freedom improves, the dominant source of noise and decoherence then shifts towards the system itself, where subsystems act as internal baths. 

In this situation, our results indicate that the MBL phase  remains robust even when coupled to an ergodic (or localized) environment in a configuration where the number of interaction terms scales with system size (the ladder configuration), provided that the coupling strength, controlled by $J_{int}$, is comparable (or smaller than) to the system's internal interaction strength, $J$. This conclusion is supported by the numerical results obtained with different total system sizes. With the steady-state staggered magnetization as a diagnostics for the preservation of information about the system's initial state, we find persisting localization that do not decay over time. Moreover, this behaviour does not depend on the initial configuration of the system and environment; memory of the system's initial state can thus be retained and extracted from local spin expectation values, regardless of the system's initial configuration and a coupling to an ergodic environment. 

As $J_{int}$ is increased, a different dynamics emerges, which depends on the system's and environment's initial configurations. We identify this behavior as resulting from the ladder being reduced into a chain of particles with both spin-0 and spin-1 degrees of freedom. The dynamics then depends on whether each dimerised ladder rung can oscillate between the two spin states under the given Hamiltonian.

Our analysis is also relevant for applications in quantum memory devices, involving the storage of information of an initial state over prolonged periods of time. Focusing on the case where the system and environment are both in the MBL phase (i.e with large disorder) and disregarding the labels ``system" and ``environment", we treat the entire ladder from the previous section as a closed memory device. If information is encoded in local spins, our results indicate the persistence of information when $J_{int}$ is comparable or lower than $J$. However, in the regime where $J_{int} \gg J$, the  inhomogeneity in energy scales in the Hamiltonian generally leads to local oscillations that destroy localization, a process dependent on the initial state of the configuration. This imposes restrictions on the relative strengths of $J_{int}$ and $J$ if these local oscillations are to be avoided.

\section{Conclusion}\label{sec:conclude}
In this work, we investigate the localization properties of MBL systems interacting with a finite environment by studying the magnetization dynamics of Heisenberg spin chains. We extensively explore the interplay between system/environment disorder strengths, geometry, initial state and system-environment coupling strengths.
We show that in most cases the system retains its localization properties despite the coupling to the environment, albeit to a reduced extent. This is supported by numerics on larger system sizes.
However, in cases where the system is strongly coupled to the environment in a ladder configuration, the eventual localization properties are highly dependent on the initial state, and could lead to either thermalization or localization. 
Our study can be experimentally implemented in multiple platforms such as trapped ion and neutral ion systems \cite{wieman1999atom,henriet2020quantum,porto2003quantum}, since the staggered magnetization used here is easily accessible in experiments. Additional numerical results using fidelity as a measure of localization and different system sizes shows that our conclusions are general. Our findings are relevant to quantum technological applications as quantum devices are increasingly miniaturized and isolated, the bath could simply be regions of quantum devices where we have limited control. 

Furthermore, our results also shed new light on quantum dynamics in the strong system-environment coupling limit with highly non-equilibrium initial states, a regime highly non-trivial for standard master equation approaches. Our findings demonstrate that the dynamics in this regime are qualitatively different from those in the weak and intermediate coupling regimes, and calls for the development of new analytical and numerical tools to investigate open quantum systems in these limits. 

\section{ACKNOWLEDGMENTS}
KLC thanks
National Research Foundation of Singapore, the Ministry of Education of Singapore, MOE grant No. MOE-T2EP50120-0019, and NSCC Project ID: 11002185 for support.

\appendix

\section{Fidelity as a state-independent diagnostic of localisation}\label{sec:fidelity}
Instead of the staggered magnetization, which we use to measure the deviation from an initial N\'eel state, more general quantifiers that are state-independent such as the fidelity or trace distance can be monitored instead. In this section, we track the fidelity instead of the staggered magnetisation as an indicator of localisation. 

The fidelity between two mixed states $\rho$ and $\sigma$ is defined as:
\begin{equation}
    F(\rho, \sigma) = (\mathrm{Tr} \sqrt{\sqrt{\rho} \sigma \sqrt{\rho}}))^2,
\end{equation}
and reduces to the squared inner product if the two states are pure.
To measure how much a quantum state has deviated from its initial state after a quantum quench, we measure the fidelity $F(\rho(t), \rho(0))$ between a time-evolved state $\rho(t)$ and its initial state $\rho(0)$, where $\rho$ can be chosen as a subsystem of a larger system.

When $\rho$ is chosen to be the pure state of the entire system-environment, $F(\rho(t), \rho(0))$ reduces to the return probability $|\braket{\psi(0)|\psi(t)}|^2$ \cite{wu2016understanding}. Following the system-environment partition in our configuration, we define $F_s \equiv F(\rho_s(t), \rho_s(0))$ and $F_e \equiv F(\rho_e(t), \rho_e(0))$ to be the system and environment fidelities respectively, where $\rho_e \equiv \mathrm{Tr}_s \rho$ and $\rho_s \equiv \mathrm{Tr}_e \rho$. We can thus track localization in individual subsystems, as we did in previous sections by studying $M_{sys}$ and $M_{env}$. The overline $\overline{F}$ similarly denotes an average over the steady-state window of the fidelity.

\begin{figure} 
         \includegraphics[width=.5\textwidth]{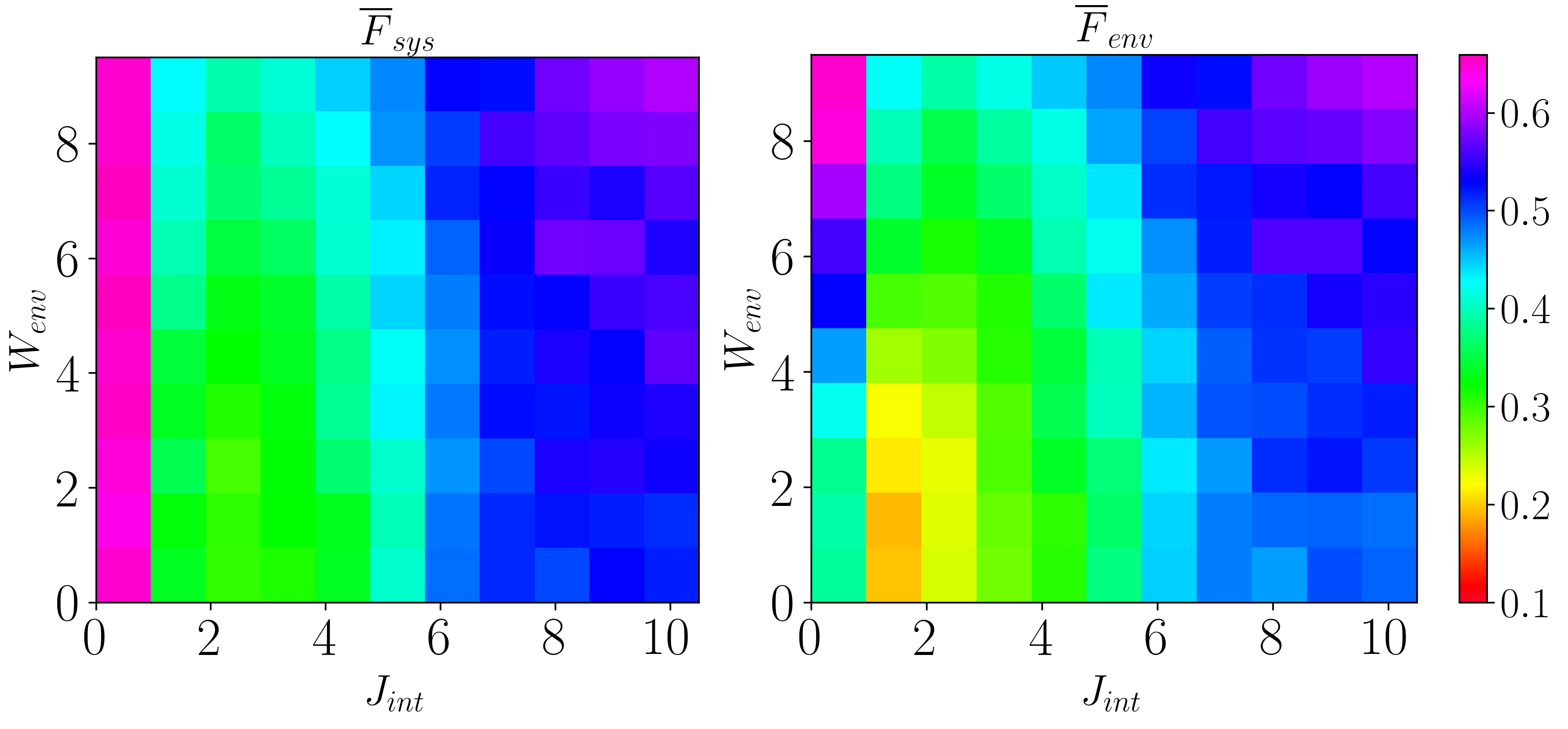}
\caption{Steady-state fidelity for the system, $\overline{F}_{sys}$, for different parameters $W_{env}$ and $J_{int}$. System-environment are in the ladder configuration, wtih an initial aligned state $\ket{\psi_A} \equiv \ket{\psi_{even}} \otimes \ket{\psi_{even}}$. The tuples $(J_{int}, W_{env})$ form a grid separated by 1 intervals. Each point is averaged over 200 disorder realisations.}
\label{fidelity}
\end{figure}

Concretely, we replicate the analysis in Section \ref{Ladder} by evolving an initially aligned strongly interacting ladder of 12 spins under the Hamiltonian $H = H_s + H_e + H^{(ladder)}$ (See Fig.~(\ref{fig:cartoon1}b)), with the initial aligned state $\ket{\psi_A}$. Computing $\overline{F}_{sys}$ and $\overline{F}_{env}$ for different parameters $J_{int} \in [0, 10]$ and $W_{env} \in [0, 9]$, we obtain Fig.~(\ref{fidelity}).

Notably, Fig.~(\ref{fidelity}) is qualitatively similar to the upper two plots of Fig.~(\ref{fig:ladder}), reproducing the characteristics described in Section \ref{Ladder} on the weak, intermediate and strong interaction regimes for the aligned initial state. $\overline{F}$ can therefore be monitored in place of the staggered magnetisation for states that are not in the N\'eel form $\ket{\uparrow \downarrow \uparrow ...}$ as a diagnostic of localization.

\section{Relation between spin chains and strongly coupled aligned ladders}\label{sec:stronglycoupled}
While the spin-1/2 ladder configuration consisting of a localised and an ergodic leg was introduced in the previous section to model the coupling of a spin chain to an external environment, Fig.~(\ref{fig:ladder_dynamics_aligned}) indicates that in the $J_{int} \gg J$ regime, rungs with initially aligned spins can encode and retain information, even when one of the legs is strongly ergodic with $W_{env}=0$. Each rung then acts as a effective spin-1 particle which form a localised 1D chain. 

In this section, we compare the retention of  information between such ladders and a 1D chain of qubits at various disorder strengths by monitoring the local magnetisation. More precisely, suppose we wish to encode and store a binary string in a chain of localised qubits that evolve under the Hamiltonian Eq.~(\ref{eq : sys hamiltonian}), with $\langle S_i \rangle > 0$ and $\langle S_i \rangle < 0$ corresponding to the two possible values of a bit. Alternatively, the results in the previous section indicates that a bit could also be encoded as the -1 and +1 states of a spin-1 particle, which we prepare as $\ket{\downarrow \downarrow}$ and $\ket{\uparrow \uparrow}$ states that evolve under Eq.~(\ref{eq:total_h}) in the ladder configuration Eq.~(\ref{eq : ladder_int}) in the $J_{int} \gg J$ regime. The binary string 1010... can then be encoded as a qubit chain in the N\'eel initial state $\ket{\uparrow \downarrow \uparrow \downarrow ...}$ and as an aligned ladder state in the form $\ket{\psi_A} \equiv \ket{\psi_{even}} \otimes \ket{\psi_{even}}$. As a measure of information retention, we again monitor the steady state staggered magnetisation, $\bar{M}$. For the qubit chain, we vary the disorder $W$ across the chain, while for the ladder, we vary both $W_{sys}$ and $W_{env}$.

\begin{figure}
     \subfloat[]{\label{equivalent_ladder}%
         \includegraphics[width=.24\textwidth]{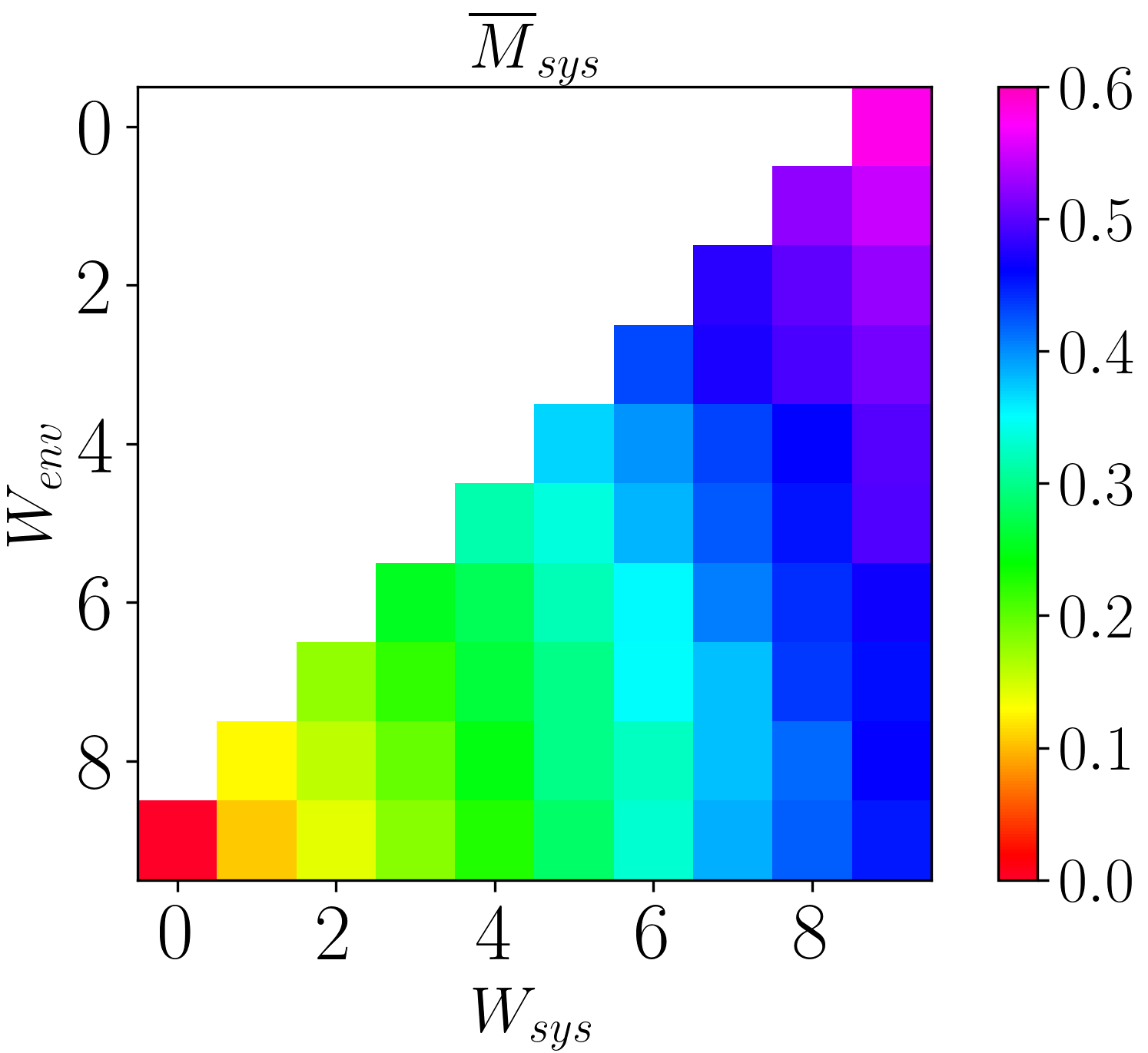}
         }
     \subfloat[]{\label{equivalent_chain}%
         \includegraphics[width=.25\textwidth]{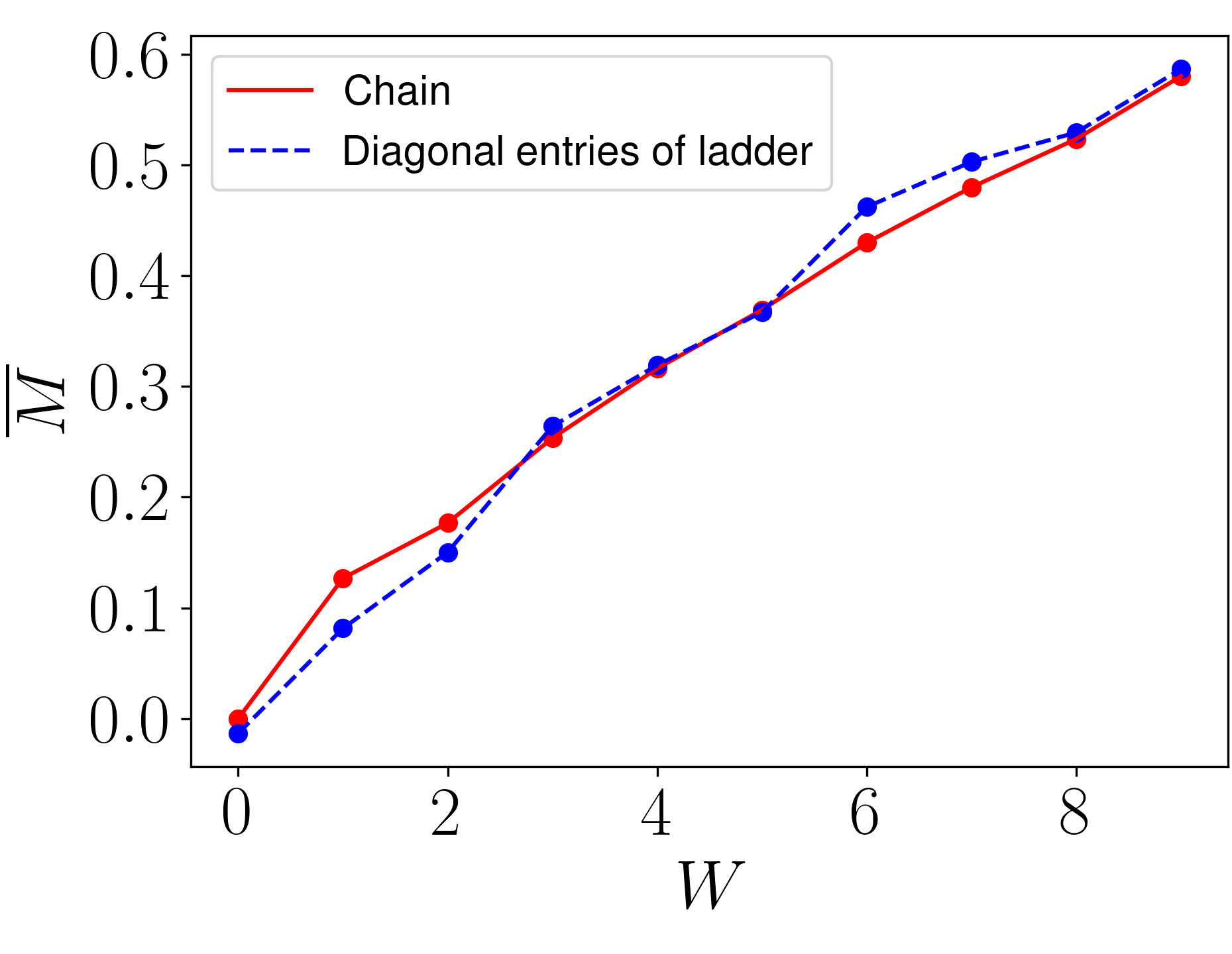}
         }
\caption{(a) Values of $\overline{M}_{sys}$ for a strongly coupled aligned ladder for different pairs of $W_{sys}$ and $W_{env}$ at $J_{int} = 20$. 
As the system and environment are equivalent and interchangeable, the matrix $\overline{M}_{sys} [W_{sys}, W_{env}]$ is symmetric, and we display only the lower triangular portion.
(b) Values of $\overline{M}$ for a 6-qubit chain for different $W$ (solid red line). For comparison, we also include the diagonal entries of the matrix $\overline{M}_{sys}[W, W]$ for the ladder considered in (a) (dotted blue line). Each point is averaged over 200 disorder realisations, and error bars are too small to be displayed.}
\end{figure}

Fig.~(\ref{equivalent_ladder}) and (\ref{equivalent_chain}) shows the steady-state staggered magnetisation $\overline{M}$ for the ladder (as $W_{sys}$ and $W_{env}$ are varied, with $J_{int} = 20$) and qubit chain (as $W$ is varied) respectively. In Fig.~(\ref{equivalent_ladder}), as the system and environment are equivalent and interchangeable, the matrix $\overline{M}_{sys} [W_{sys}, W_{env}]$ is symmetric, and we display only the lower triangular portion. Moreover, $\overline{M}_{sys} \approx \overline{M}_{env}$ (As $J \ll J_{int}$; see \ref{strong_int_reg}), so we display only $\overline{M}_{sys}$.

From Fig.~(\ref{equivalent_chain}), we observe that a strongly interacting aligned ladder with $W_{sys} = W_{env} = W$ (dotted blue line) can reproduce the localisation properties of a 6-qubit chain with disorder $W$ (red line). That is, the same amount of information can be encoded in the +1 and -1 states of a spin-1 particle or in the +1/2 and -1/2 states of a spin-1/2 particle with the choice $W_{sys} = W_{env} = W$; information loss to the 0 state of the spin-1 mode and the spin-0 mode appear negligible. We also observe from Fig.~(\ref{equivalent_ladder}) that different choices of $W_{sys}$ and $W_{env}$ can lead to the same staggered magnetisation (there are multiple regions with the same colour in Fig.~(\ref{equivalent_ladder})).

\section{Diagonal ensemble}\label{sec:diagonal_ensemble}
Here, we briefly describe how infinite-time averages of observables can be obtained by considering the diagonal ensemble \cite{rigol2008thermalization}, as we did in Fig.~(\ref{fig:junc_dynamics}).
For an initial state $\ket{\psi(0)} = \sum_i C_i \ket{\phi_i}$ expanded in the eigenbasis $\{\ket{\phi_i}\}$ of a Hamiltonian $H$, its evolution under the same Hamiltonian is:
\begin{equation}
    \ket{\psi(t)} = \sum_i C_i e^{-i E_i t} \ket{\phi_i}.
\end{equation}
The expectation value of any observable $\hat{O}$ then evolves as:
\begin{equation}
    \left\langle \psi(t) \middle| \hat{O} \middle| \psi (t) \right\rangle = \sum_{i,j} C_i^* C_j e^{-i (E_i - E_j) t} \left\langle \phi_i \middle| \hat{O} \middle| \phi_j \right\rangle.
\end{equation}
Taking the infinite time average of this quantity, the off-diagonal terms of $O$ are oscillatory and hence average to zero, leaving the diagonal terms remaining:
\begin{equation}
    \overline{\left\langle \psi(t) \middle| \hat{O} \middle| \psi (t) \right\rangle} = \sum_i |C_i|^2 \left\langle \phi_i \middle| \hat{O} \middle| \phi_i \right\rangle.
\end{equation}
We can thus compute infinite-time averages of observables such as the staggered magnetisation in Eq.~(\ref{eq : stag_mag_expt}) by diagonalizing $H$ to obtain $\{\ket{\phi_i}\}$.
Alternatively, one can take the time average of the observable after evolving the system for a long time as in Eq.~(\ref{eq : ss_stagmag}). Fig.~(\ref{fig:junc_dynamics}) shows expectation values obtained from the two approaches for the staggered magnetisation and local spin operators. We find good agreement between the two approaches, and expect that evolving and averaging over a longer period of time will yield convergence to the diagonal ensemble values.

\bibliography{references}

\end{document}